\documentclass[aps,prl,twocolumn,showpacs,superscriptaddress,groupedaddress]{revtex4}  
\usepackage{graphicx}  
\usepackage{dcolumn}   
\usepackage{bm}        
\usepackage{amssymb}   

\newcommand{\pD}[2]{\frac{\partial #2}{\partial #1}}

\newcommand{\D}[2]{\frac{{\rm d} #2}{{\rm d} #1}}

\newcommand\bb[1]{\mbox{\boldmath{$#1$}}}
\newcommand\grad{\bb{\nabla}}
\newcommand\bcdot{\bb{\cdot}}
\newcommand\btimes{\bb{\times}}

\newcommand{\ex}{\hat{\bb{x}}}
\newcommand{\ey}{\hat{\bb{y}}}

\begin{document}

\title{Firehose and Mirror Instabilities in a Collisionless Shearing Plasma}
\author{Matthew W.~Kunz}\email{mkunz@princeton.edu} \affiliation{Department of Astrophysical Sciences, Princeton University, 4 Ivy Lane, Princeton, NJ 08544, USA}
\author{Alexander A.~Schekochihin} \affiliation{Rudolf Peierls Centre for Theoretical Physics, University of Oxford, 1 Keble Road, Oxford OX1 3HQ, UK} \affiliation{Merton College, Merton St, Oxford OX1 4JD, UK}
\author{James M. Stone} \affiliation{Department of Astrophysical Sciences, Princeton University, 4 Ivy Lane, Princeton, NJ 08544, USA}

\date{\today}

%
%
\begin{abstract}
Hybrid-kinetic numerical simulations of firehose and mirror instabilities in a collisionless plasma are performed in which pressure anisotropy is driven as the magnetic field is changed by a persistent linear shear $S$. For a decreasing field, it is found that mostly oblique firehose fluctuations grow at ion Larmor scales and saturate with energies $\propto$$S^{1/2}$; the pressure anisotropy is pinned at the stability threshold by  particle scattering off microscale fluctuations. In contrast, nonlinear mirror fluctuations are large compared to the ion Larmor scale and grow secularly in time; marginality is maintained by an increasing population of resonant particles trapped in magnetic mirrors. After one shear time, saturated order-unity magnetic mirrors are formed and particles scatter off their sharp edges. Both instabilities drive sub-ion-Larmor--scale fluctuations, which appear to be kinetic-Alfv\'{e}n-wave turbulence. Our results impact theories of momentum and heat transport in astrophysical and space plasmas, in which the stretching of a magnetic field by shear is a generic process.
\end{abstract}

\maketitle

%
%
\paragraph{Introduction.}
Describing the large-scale behavior of weakly collisional magnetized plasmas, such as the solar wind, hot accretion flows, or the intracluster medium (ICM) of galaxy clusters, necessitates a detailed understanding of the kinetic-scale physics governing the dynamics of magnetic fields and the transport of momentum and heat. This physics is complicated by the fact that such plasmas are expected to exhibit particle distribution functions with unequal thermal pressures in the directions parallel ($||$) and perpendicular ($\perp$)  to the local magnetic field \cite{msrmpn82,sc06,shqs06}. This pressure anisotropy can trigger fast micro-scale instabilities \cite{rosenbluth56,ckw58,parker58,vs58,barnes66,hasegawa69}, whose growth and saturation impact the structure of the magnetic field and the effective viscosity of the plasma. While solar-wind observations suggest that these instabilities are effective at regulating the pressure anisotropy to marginally stable levels \cite{gsss01,klg02,htkl06,matteini07,bkhqss09,mhglvn13}, it is not known how this is achieved.

We address this question with nonlinear numerical simulations of the firehose and mirror instabilities. We leverage the universal physics at play in turbulent $\beta \gg 1$ astrophysical plasmas such as the ICM \cite{sckhs05,kscbs11} and Galactic accretion flows \cite{qdh02,rqss12}---magnetic field being changed by velocity shear, coupled with adiabatic invariance---to drive self-consistently a pressure anisotropy beyond the instability thresholds. Our setup represents a local patch of a turbulent velocity field, in which the magnetic field is sheared and its strength changed on a timescale much longer than that on which the unstable fluctuations grow. This approach is complementary to expanding-box models of the $\beta \sim 1$ solar wind \cite{gv96} used to drive firehose \cite{mlhv06,ht08} and mirror/ion-cyclotron \cite{ht05} instabilities.

%
%
\paragraph{Hybrid-kinetic equations in the shearing sheet.}
A non-relativistic, quasi-neutral, collisionless plasma of electrons (mass $m_{\rm e}$, charge $-e$) and ions (mass $m_{\rm i}$, charge $Ze$) is embedded in a linear shear flow, $\bb{u}_0 = - S x \ey$, in $(x,y,z)$ Cartesian coordinates. In a frame co-moving with the shear flow, the equations governing the evolution of the ion distribution function $f_{\rm i} (t, \bb{r}, \bb{v})$ and the magnetic field $\bb{B}$ are, respectively, the Vlasov equation
\begin{equation}\label{eqn:vlasov}
\D{t}{f_{\rm i}} + \bb{v} \bcdot \grad f_{\rm i} + \left[ \frac{Ze}{m_{\rm i}} \left( \bb{E}' + \frac{\bb{v}}{c} \btimes \bb{B} \right) + S v_x \ey \right] \! \bcdot \pD{\bb{v}}{f_{\rm i}} = 0 
\end{equation}
and Faraday's law
\begin{equation}\label{eqn:induction}
\D{t}{\bb{B}} = - c \grad \btimes \bb{E}' - S B_x \ey ,
\end{equation}
where ${\rm d} / {\rm d} t \equiv \partial / \partial t - S x \, \partial / \partial y$. The electric field,
\begin{equation}\label{eqn:efield}
\bb{E}' = - \frac{\bb{u}_{\rm i} \btimes \bb{B}}{c} + \frac{( \grad \btimes \bb{B} ) \btimes \bb{B}}{4\pi Z e n_{\rm i}} - \frac{T_{\rm e} \grad n_{\rm i}}{e n_{\rm i}} ,
\end{equation}
is obtained by expanding the electron momentum equation in $( m_{\rm e} / m_{\rm i} )^{1/2}$, enforcing quasi-neutrality
\begin{equation}\label{eqn:quasineutrality}
n_{\rm e} = Z n_{\rm i} \equiv Z \! \int {\rm d}^3 \bb{v} \, f_{\rm i} ,
\end{equation}
assuming isothermal electrons, and using Amp\`{e}re's law to solve for the mean velocity of the electrons
\begin{equation}\label{eqn:ampere}
\bb{u}_{\rm e} = \bb{u}_{\rm i} - \frac{\bb{j}}{Z e n_{\rm i}} \equiv  \frac{1}{n_{\rm i}} \int {\rm d}^3 \bb{v} \, \bb{v} f_{\rm i} - \frac{ c \grad \btimes \bb{B} }{4 \pi Z e n_{\rm i}} 
\end{equation}
in terms of the mean velocity of the ions $\bb{u}_{\rm i}$ and the current density $\bb{j}$ \cite{rsrc11,rsc14}. This constitutes the ``hybrid'' description of kinetic ions and fluid electrons \cite{bcch78,hn78}.
%
%
\paragraph{Adiabatic invariance and pressure anisotropy.}
The final terms in Eqs.~(\ref{eqn:vlasov}) and (\ref{eqn:induction}) represent the stretching of the phase-space density and the magnetic field in the $y$-direction by the shear flow. Conservation of the first adiabatic invariant $\mu \equiv m_{\rm i} v^2_\perp / 2B$ then renders $f_{\rm i}$ anisotropic with respect to the magnetic field. If $\bb{E}' = 0$, the ratio of the perpendicular and parallel pressures is
\begin{equation}\label{eqn:paniso}
\frac{p_\perp}{p_{||}} \equiv \frac{ \int {\rm d}^3 \bb{v} \, \mu B \, f_{\rm i}}{\int {\rm d}^3 \bb{v} \, m_{\rm i} v^2_{||} \, f_{\rm i} } = \left[ 1 - 2 \frac{B_x B_{y0}}{B^2_0} S t + \frac{B^2_x}{B^2_0} ( S t )^2 \right]^{3/2} ,
\end{equation}
where the subscript `$0$' denotes initial values \cite{cgl56}.
%
%
\paragraph{Method of solution.}
We solve Eqns.~(\ref{eqn:vlasov})--(\ref{eqn:ampere}) using the second-order--accurate particle-in-cell code {\sc Pegasus} \cite{ksb14}. We normalize magnetic field to $B_0$, velocity to the initial Alfv\'{e}n speed $v_{\rm A0} \equiv B_0 / \sqrt{4\pi m_{\rm i} n_{\rm i0}}$, time to the inverse of the initial ion gyrofrequency $\Omega_{\rm i0} \equiv Z e B_0 / m_{\rm i} c$, and distance to the initial ion skin depth $d_{\rm i0} \equiv v_{\rm A0} / \Omega_{\rm i0}$. The ion Larmor radius $\rho_{\rm i} = \beta^{1/2}$, where $\beta \equiv 8 \pi n_{\rm i} T_{\rm i} / B^2$. $N_{\rm p}$ particles are drawn from a Maxwell distribution with $\beta_0 = 200$ and placed on a 2D grid $N_x$$\times$$N_y = 1152^2$ cells spanning $L_x$$\times$$L_y = 1152^2$. The electrons are Maxwellian and gyrotropic with $T_{\rm i} / Z T_{\rm e} = 1$. A $\delta f$ method reduces the impact of discrete-particle noise on the moments of $f_{\rm i}$ \cite{pl93,hk94}. Orbital advection updates the particle positions and magnetic field due to the background shear \cite{sg10}. The boundary conditions are shearing-periodic: $f(x,y) = f(x \pm L_x , y \mp S L_x t)$. We scan $S = (1,3,10,30) \times 10^{-4}$. These parameters guarantee a healthy scale separation between the grid scale, the ion Larmor radius, the wavelengths of the instabilities, and the box size. In what follows, $\langle \cdot \rangle$ denotes a spatial average over all cells.

\begin{figure}
\centering
\includegraphics[width=8.5cm,clip]{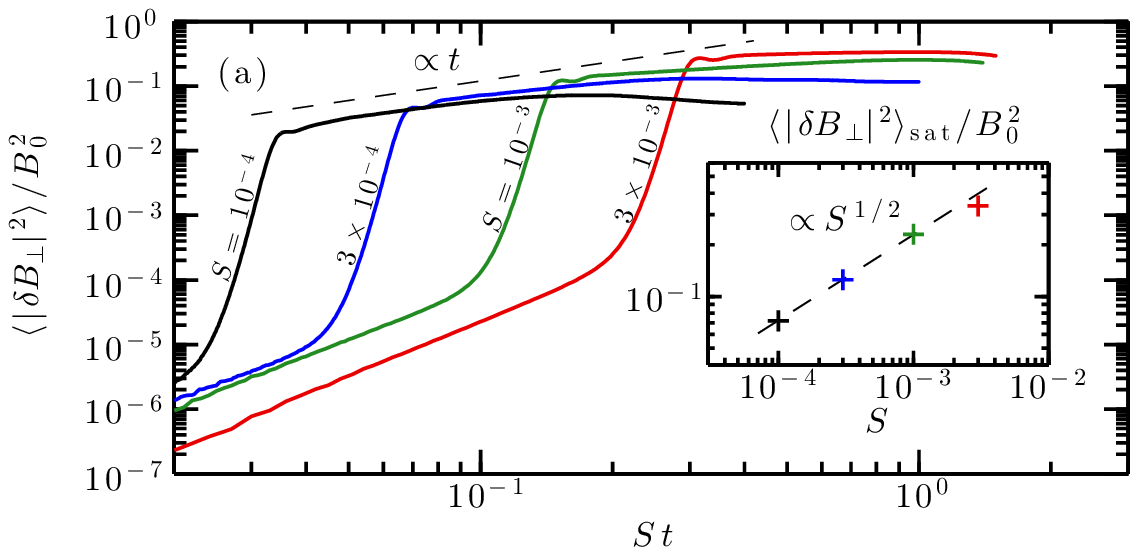}
\newline
\includegraphics[width=8.5cm,clip]{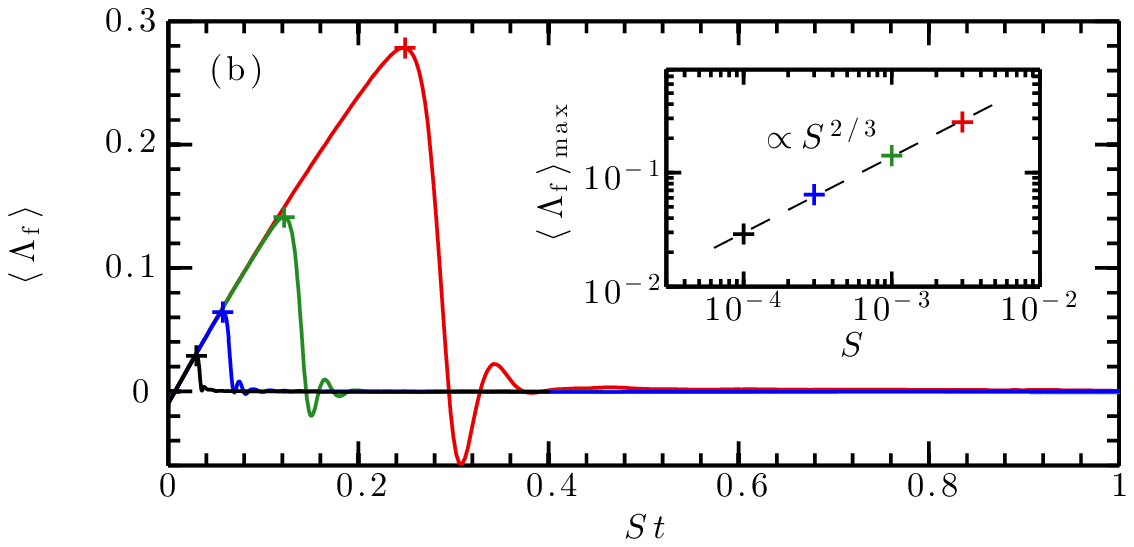}
\newline
\vspace{-0.5cm}
\caption{Evolution of firehose instability. (a) Energy in perpendicular magnetic fluctuations, $\langle | \delta \bb{B}_\perp |^2 \rangle$, whose saturated value $\propto$$S^{1/2}$ (inset). (b) Firehose stability parameter, $\langle \Lambda_{\rm f} \rangle$, whose maximum value $\propto$$S^{2/3}$ (inset; see text for explanation).}
\label{fig:fhs-boxavg}
\end{figure}

\begin{figure}
\centering
\includegraphics[width=8.5cm,clip]{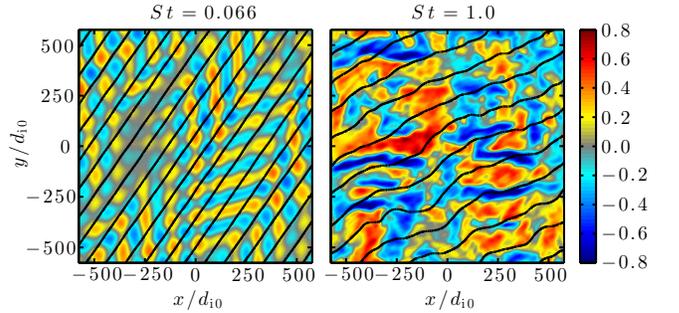}
\vspace{-0.2cm}
\caption{Spatial structure of the firehose instability with $S = 3\times 10^{-4}$. $\delta B_z / B_0$ (color) and magnetic-field lines are shown in the linear ({\it left}) and saturated ({\it right}) regimes.}
\label{fig:fhs}
\end{figure}

%
%
\paragraph{Firehose instability.}
We choose $N_{\rm p} = 1024 N_x N_y$ and set $\bb{B}_0 = (2 \ex + 3 \ey ) / \sqrt{13}$, so that $\langle B_{y} \rangle = \langle B_{x} \rangle$ at $St = 1/2$. As $B$ decreases, adiabatic invariance drives $p_\perp / p_{||} < 1$ (Eq.~\ref{eqn:paniso}), with plasma becoming firehose unstable when $\Lambda_{\rm f} \equiv 1 - p_\perp / p_{||} - 2 / \beta_{||} > 0$. Exponentially growing, Alfv\'{e}nically polarized ($|\delta \bb{B}_\perp| \gg \delta B_{||}$), oblique modes with growth rate $\gamma \simeq k_{||} \rho_{\rm i} ( \Lambda_{\rm f} / 2 )^{1/2}$ and $k_{||} \rho_{\rm i} \approx k_\perp \rho_{\rm i} \approx 0.4$ then appear (Fig.~\ref{fig:fhs-boxavg}a; cf.~\cite{ywa93,hm00}). Fig.~\ref{fig:fhs} shows their spatial structure. $\Lambda_{\rm f}$ continues to grow, driven by shear ($\Lambda_{\rm f} \sim St$; Fig.~\ref{fig:fhs-boxavg}b), until the perturbations become large enough to reduce the pressure anisotropy to its marginally stable value ($\Lambda_{\rm f} \rightarrow 0$).

\begin{figure}
\centering
\includegraphics[width=8.5cm,clip]{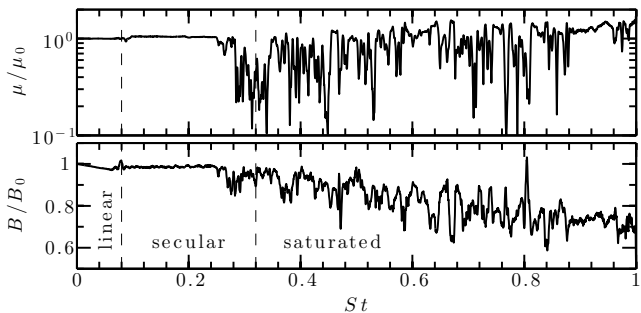}
\vspace{-0.2cm}
\caption{Evolution of $\mu$ and $B$ for a representative particle in the firehose simulation with $S = 3 \times 10^{-4}$.}
\label{fig:fhs-scattering}
\end{figure}

It has been proposed \cite{sckrh08,rsrc11} that they do this by canceling the rate of change of the mean field: $(1/2) \, {\rm d} \langle | \delta \bb{B}_\perp |^2 \rangle / {\rm d} t \approx - {\rm d} \ln | \langle \bb{B} \rangle | / {\rm d} t \sim S$, giving rise to secular evolution, $\langle | \delta \bb{B}_\perp |^2 \rangle \sim S t$. Matching $\gamma \sim \Lambda^{1/2}_{\rm f} \sim (St)^{1/2}$ with the rate of growth in the secular phase ($\gamma \sim 1/t$), we find $\langle | \delta \bb{B}_\perp |^2 \rangle \sim St \sim \Lambda_{\rm f} \sim S^{2/3}$ at the transition from linear to nonlinear evolution (cf.~\cite{ss64,qs96}; ``quasi-linear saturation"). This scenario is indeed what we observe: the evolution of $\langle | \delta \bb{B}_\perp |^2 \rangle$ and $\Lambda_{\rm f}$ is shown in Fig.~\ref{fig:fhs-boxavg}; note $\langle \Lambda_{\rm f} \rangle_{\rm max} \propto S^{2/3}$ (inset in Fig.~\ref{fig:fhs-boxavg}b). To test the idea \cite{sckrh08,rsrc11} that, during the secular phase, the average $B$ seen by particles streaming along the field is constant, we plot in Fig.~\ref{fig:fhs-scattering} a representative particle's $\mu$ and $B$ (evaluated at the particle's position) for $S = 3 \times 10^{-4}$. During the secular phase, the particle nearly conserves $\mu$ and $B \simeq {\rm const}$ along its trajectory, as expected. 

%
%

%
%

\begin{figure}
\centering
\includegraphics[width=8.5cm,clip]{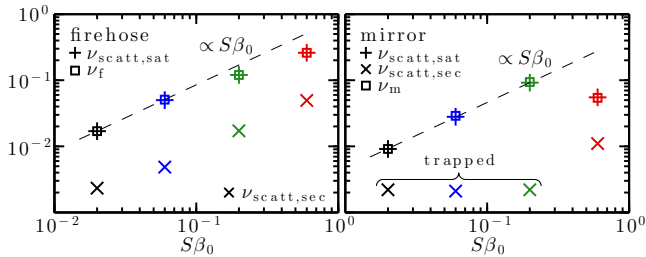}
\vspace{-0.2cm}
\caption{Mean scattering rate $\nu_{\rm scatt}$ for ({\it left}) firehose and ({\it right}) mirror in the secular (crosses) and saturated (plus signs) phases versus $S \beta_0$. The collision rates required to maintain marginal stability in the saturated phase, $\nu_{\rm f}$ and $\nu_{\rm m}$ respectively, are shown for comparison. See text for definitions.}
\label{fig:nuscat}
\end{figure}

However, this secular growth is not sustainable: the magnetic fluctuation energy saturates at a low level $\propto$$S^{1/2}$ (inset of Fig.~\ref{fig:fhs-boxavg}a) in a state of firehose turbulence. During this saturated state, particles scatter off fluctuations with $k_{||} \rho_{\rm i} \sim 1$, $\mu$ conservation is broken, and $B$ decreases at a rate approaching $-{\rm d} \ln | \langle \bb{B} \rangle | / {\rm d} t \sim S$ (Fig.~\ref{fig:fhs-scattering}). The production of pressure anisotropy is no longer adiabatically tied to the rate of change of the magnetic field and marginality ($\Lambda_{\rm f} \simeq 0$) is maintained independently of $S$ via anomalous particle scattering. We calculate the mean scattering rate $\nu_{\rm scatt}$ by tracking 4096 randomly selected particles, constructing a distribution of times taken by each to change its $\mu$ by a factor of ${\rm e}$, and taking the width of the resulting exponential function to be $\nu^{-1}_{\rm scatt}$. In a collisional, incompressible plasma without heat flows, the pressure anisotropy would be $p_\perp / p_{||} - 1 = (3 / \nu) ( {\rm d} \ln | \langle \bb{B} \rangle | / {\rm d} t )$, where $\nu$ is collision rate \cite{braginskii65,rsrc11}. The effective scattering rate needed to maintain $\Lambda_{\rm f} = 0$ at saturation would then be $\nu_{\rm f} \equiv -3 (\beta_{||,{\rm sat}} /2) ( {\rm d} \ln | \langle \bb{B} \rangle | / {\rm d}t )_{\rm sat} \sim S \beta$. Remarkably, we find $\nu_{\rm scatt} \simeq \nu_{\rm f}$ in the saturated state (Fig.~\ref{fig:nuscat}).

%
%

%
%
%

\begin{figure}
\centering
\includegraphics[width=8.5cm,clip]{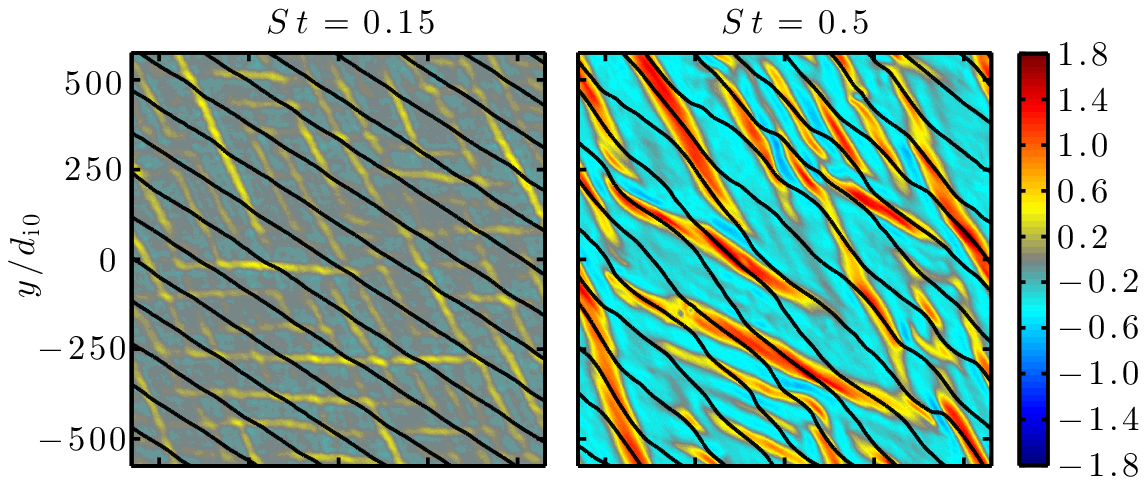}
\newline
\includegraphics[width=8.5cm,clip]{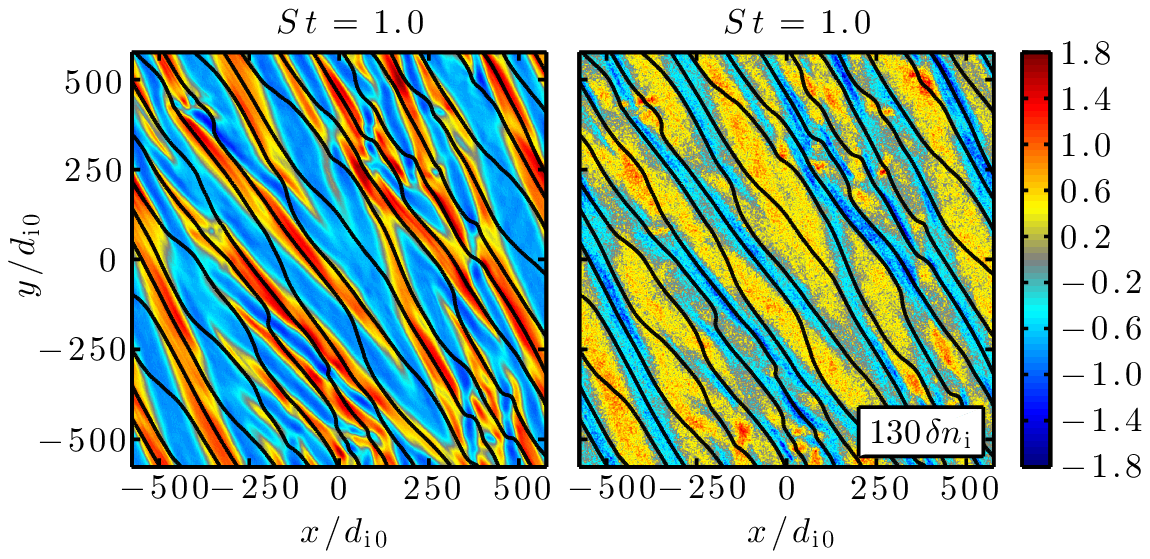}
\newline
\vspace{-0.5cm}
\caption{Spatial structure of the mirror instability with $S = 3\times 10^{-4}$. $\delta B_{||} / B_0$ and (last panel) re-scaled $\delta n_{\rm i} / n_{\rm i0}$ are shown (color) with magnetic-field lines in the shearing plane.}
\label{fig:mrs}
\end{figure}

%
%
\paragraph{Mirror instability.}
We choose $N_{\rm p} = 625 N_x N_y$ and set $\bb{B}_0 = (2 \ex - \ey ) / \sqrt{5}$, so that $\langle B_{y} \rangle = - \langle B_{x} \rangle$ at $St = 1/2$. As $B$ increases, adiabatic invariance drives $p_\perp / p_{||} > 1$ (Eq.~\ref{eqn:paniso}), with plasma becoming mirror unstable when $\Lambda_{\rm m} \equiv p_\perp / p_{||}  - 1 - 1/\beta_\perp > 0$ \footnote{Technically, this instability parameter is for cold electrons \cite{hellinger07}, but it is close enough to reality for simplicity to outweigh precision in our treatment.}. Near threshold, linearly growing perturbations have $\gamma \sim \Lambda^2_{\rm m}$, $k_{||} \rho_{\rm i} \sim \Lambda_{\rm m}$, and $k_\perp \rho_{\rm i} \sim \Lambda_{\rm m}^{1/2}$ \citep{hellinger07}---they grow slower than the firehose, are more elongated in the magnetic-field direction, and have $\delta B_{||} \gg | \delta \bb{B}_\perp |$. Fig.~\ref{fig:mrs} shows their spatial structure.

\begin{figure}
\centering
\includegraphics[width=8.5cm,clip]{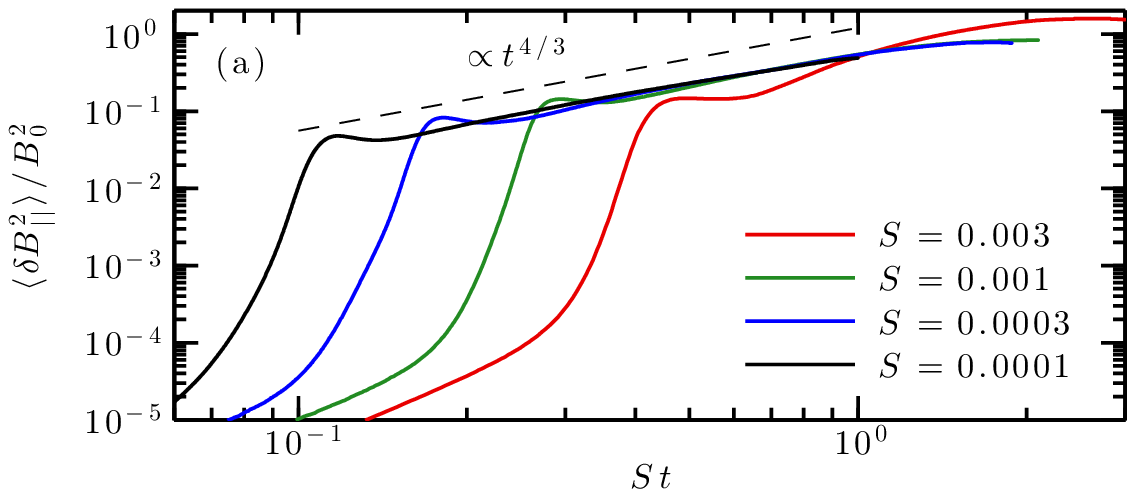}
\newline
\includegraphics[width=8.5cm,clip]{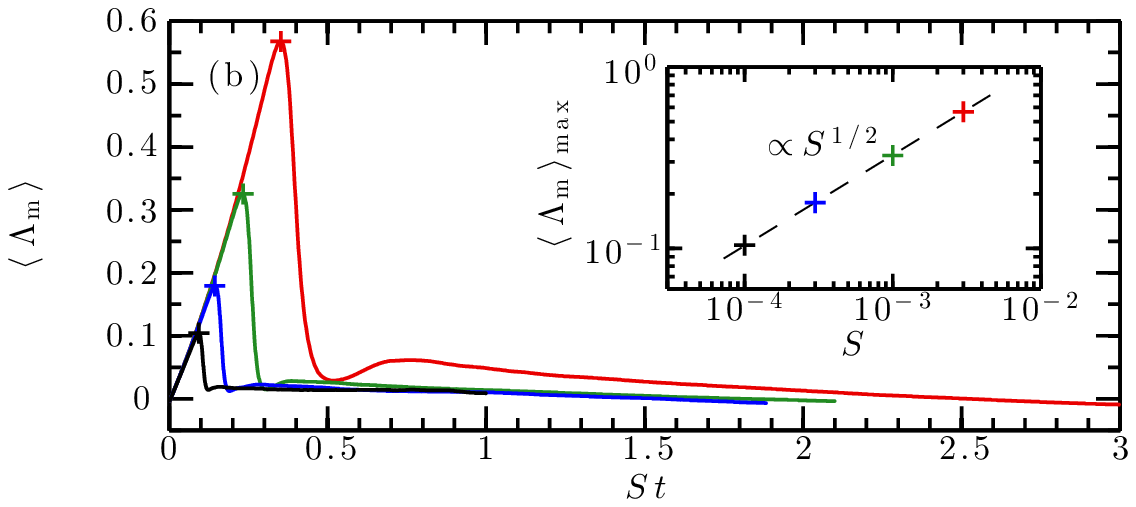}
\newline
\vspace{-0.5cm}
\caption{Evolution of mirror instability versus $S$. (a) Energy in parallel fluctuations of the magnetic field, $\langle \delta B^2_{||} \rangle$. (b) Mirror stability parameter, $\langle \Lambda_{\rm m} \rangle$, whose maximum value $\propto$$S^{1/2}$.}
\label{fig:mrs-boxavg}
\end{figure}

The saturation scenario is analogous to the firehose: $\Lambda_{\rm m}$ continues growing (Fig.~\ref{fig:mrs-boxavg}b) until the mirror perturbations are large enough to drive $\Lambda_{\rm m} \rightarrow 0$, at which point the perturbations' exponential growth gives way to secular evolution with $\langle \delta B^2_{||} \rangle \propto t^{4/3}$ (Fig.~\ref{fig:mrs-boxavg}a, discussed below). As $\Lambda_{\rm m} \rightarrow 0$, the dominant modes shift to longer wavelengths ($k_{||} \rho_{\rm i} \ll 1$) and become more elongated in the mean-field direction. Excepting the (non-asymptotic) $S = 10^{-3}$ case, this secular phase appears to be universal, lasting until $\delta B / B_0 \sim 1$ at $S t \gtrsim 1$, independently of $S$. The final saturation is caused by particle scattering off sharp ($\delta B / B_0 \sim 1$, $k_{||} \rho_{\rm i} \sim 1$) bends in the magnetic field, which occur at the boundaries of the magnetic mirrors.

As foreseen by \cite{sk93,ks96,pantellini98}, trapped particles play a crucial role in the nonlinear evolution. Following \cite{sckrh08,rsc14}, we expect the pressure anisotropy to be pinned at marginal by an increasing fraction ($\sim$$|\delta B_{||}|^{1/2}$) of particles becoming trapped in magnetic mirrors, thereby sampling regions where the increase of the mean field is compensated by the decrease in the perturbed field, viz.~$ - {\rm d} \overline{\delta B_{||}} / {\rm d} t \sim {\rm d} \langle | \delta B_{||} |^{3/2} \rangle / {\rm d} t \sim {\rm d} \ln | \langle \bb{B} \rangle | / {\rm d} t \sim S$, where the overbar denotes averaging along particle trajectories (i.e.~bounce-averaging for trapped particles). It follows that $\langle \delta B^2_{||} \rangle \sim ( S t )^{4/3}$, as is indeed seen in Fig.~\ref{fig:mrs-boxavg}a.

%
%
%

%
%

\begin{figure}
\centering
\includegraphics[width=8.5cm,clip]{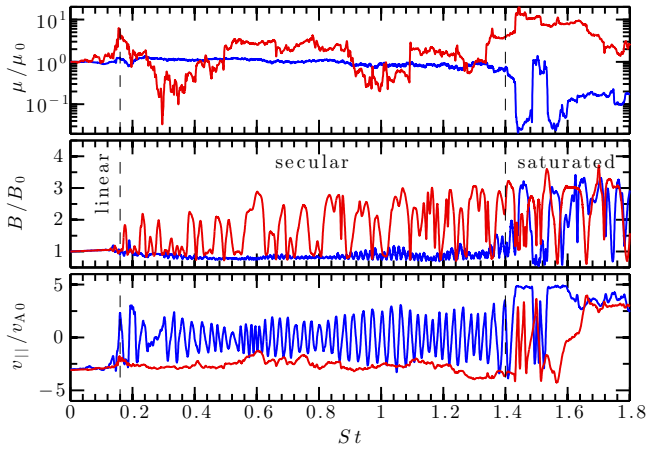}
\vspace{-0.1in}
\caption{Evolution of $\mu$, $B$, and $v_{||}$ (evaluated at particle position) for representative passing (red) and trapped (blue) particles in the mirror simulation with $S = 3 \times 10^{-4}$.}
\label{fig:mrs-scattering}
\end{figure}

Fig.~\ref{fig:mrs-scattering} displays $\mu$, $B$, and $v_{||}$ for representative passing and trapped particles in the simulation with $S = 3 \times 10^{-4}$. In the linear phase, both particles conserve $\mu$ very well. During the secular phase ($St \simeq 0.2$--$1.4$), one of the particles becomes trapped and bounces while nearly conserving $\mu$; $B \simeq {\rm const}$ along its path, despite the growing mean field. The other remains passing, with $\overline{\delta B_{||}} \approx 0$. At the end of the secular phase, the trapped particle scatters out of the mirror and becomes passing.

The mean scattering rates $\nu_{\rm scatt}$ are different for the trapped and passing populations. During the secular phase, the trapped particles ($\sim$$70\%$ towards the end of the secular phase \footnote{This is consistent with the fraction of trapped particles being $f_{\rm T} = ( 1 - B_{\rm min} / B_{\rm max} )^{1/2}$. For $B_{\rm max} / B_{\rm min} \simeq 1.8$ near the end of the secular phase (see Fig.~\ref{fig:mrs}), $f_{\rm T} \simeq 0.66$.}) have $\nu_{\rm scatt} \approx 0.002$ (Fig.~\ref{fig:nuscat}), while the passing particles have $\nu_{\rm scatt} \approx 0.03$. Excepting the $S = 10^{-3}$ case, these values are independent of $S$, indicating that particle scattering is irrelevant for $St \lesssim 1$ and $S \ll 1$. At saturation ($St \gtrsim 1$), the percentage of trapped particles drops to $\sim$$30\%$ (with $\nu_{\rm scatt} \approx 0.004$) and the total $\nu_{\rm scatt} \simeq \nu_{\rm m}$, where $\nu_{\rm m} \equiv 3 \beta_{\perp,\rm sat} ( {\rm d} \ln | \langle \bb{B} \rangle | / {\rm d} t )_{\rm sat}$ is the collisionality required to maintain $\Lambda_{\rm m}  = 0$ at saturation (by the same argument as in the firehose discussion).

%
%
%

\begin{figure}
 \centering
 \includegraphics[width=8.5cm,clip]{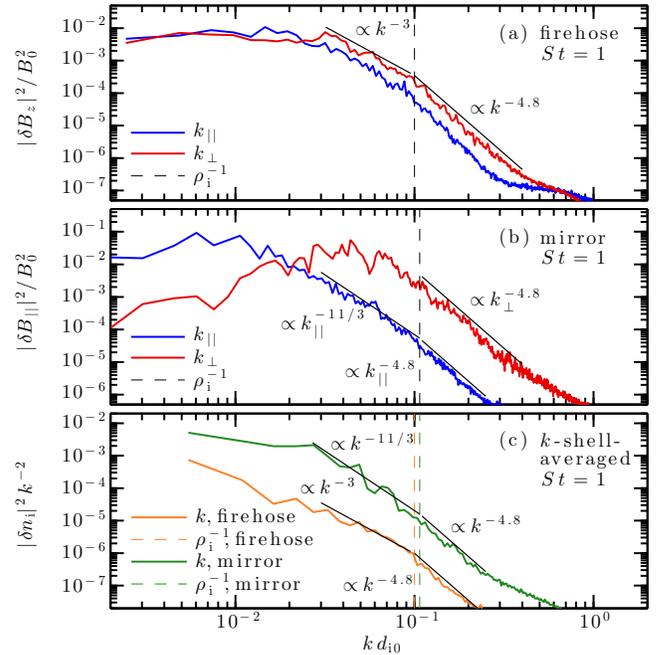}
 \vspace{-0.1in}
 \caption{1D magnetic fluctuation spectra for (a) firehose and (b) mirror versus $k_{||}$ and $k_\perp$, and (c) $k$-shell-averaged density fluctuation spectra for firehose and mirror versus $k$, all in the saturated state ($St = 1$) of the $S = 3 \times 10^{-4}$ simulations.}
 \label{fig:spectra}
 \end{figure}

%
%
\paragraph{Firehose- and mirror-driven turbulence.}
The saturated state of both instabilities is characterized by super-Larmor-scale driving and sub-Larmor-scale fluctuations. Fig.~\ref{fig:spectra} shows 1D magnetic fluctuation spectra for firehose and mirror at saturation versus $k_{||}$ and $k_\perp$ for $S = 3 \times 10^{-4}$. Energy is injected at successively larger scales as marginality is approached \cite[cf.][]{qs96,mlhv06} and several power laws are established. Firehose modes with $k \rho_{\rm i} < 1$ satisfy $| \delta B_{z,k} |^2 \propto k^{-3}$, a spectrum reminiscent of that predicted for parallel-firehose turbulence \cite{rsrc11}. Mirror modes with $k \rho_{\rm i} < 1$ satisfy $| \delta B_{||,k_{||}} |^2 \propto k^{-11/3}_{||}$. This scaling is obtained by an argument analogous to that proposed in \cite{rsrc11}: seek a power-law spectrum, $| \delta B_{||,k_{||}} |^2 \sim k^{-\alpha}_{||}$; estimate $\gamma_{\rm peak} \sim \Lambda^2_{\rm m} \sim 1/t$ and $k_{||,{\rm peak}} \sim \Lambda_{\rm m} \sim 1/t^{1/2}$ for the energy-containing mode in the secular phase; recall $\sum_{k_{||}} | \delta B_{||,k_{||}} |^2 \sim (S t)^{4/3}$; and demand that this be consistent with $\sum_{k_{||}} | \delta B_{||,k_{||}}|^2 \sim k_{||,{\rm peak}}^{1-\alpha} \sim t^{-(1-\alpha)/2}$. This procedure yields $\alpha = 11/3$. Finally, the $k$-shell-averaged density fluctuation spectra (Fig.~\ref{fig:spectra}c) follows $|\delta B_{||}|^2$, as expected for pressure-balanced mirrors.

Both spectra indicate that energy is removed at sub-Larmor scales by what appears to be a turbulent cascade, whose spectral slope and polarization of the fluctuations ($\delta n_{\rm i} \sim \beta^{-1}\, \delta B_{||}$ \cite{scdhhqt09,bhxp13}) approximately matches observations of KAW turbulence in gyrokinetic simulations \cite{htdqsnt11} and the solar wind \cite{sgbcr10,aslmmsr09,cbxp13}, as well as of ``mirror turbulence" in the magnetosheath \cite{sbrcpb06}. This marks the first time in a simulation of mirror or firehose turbulence that a KAW cascade has been observed. Nevertheless, we caution that our simulations were performed in 2D; a proper study of this cascade requires 3D geometry \cite{scdhhqt09,htdqsnt11,bp12}.
 
 %
 %
%

%
%
\paragraph{Summary.}
We have presented numerical simulations of firehose and mirror instabilities driven by a changing magnetic field in a local shear flow. Both instabilities start in the linear regime with exponential growth, a process that is well understood analytically. The theoretical expectation, that after linear saturation the growth becomes secular as the pressure anisotropy is persistently driven \cite{sckrh08,rsrc11,rsc14}, is borne out by our simulations. For the firehose, the marginal state is initially achieved via $\mu$-conserving changes in the magnetic field, but is subsequently maintained (independent of $S$) by particle scattering off $k_{||} \rho_{\rm i} \sim 1$ fluctuations. For the mirror, marginal stability is achieved and maintained during the secular phase by particle trapping in magnetic mirrors. Saturation occurs once $\delta B / B_0 \sim 1$ at $St \gtrsim 1$ via particle scattering off the sharp ends of the mirrors. For both instabilities, the mean scattering rate at saturation adjusts to maintain marginal stability, effectively reducing the viscosity to $v^2_{\rm th} / \nu_{\rm scatt} \sim v^2_{\rm A,sat} / S$.

%
%
\begin{acknowledgments}
Support for MWK was provided by NASA through Einstein Postdoctoral Fellowship Award Number PF1-120084, issued by the {\it Chandra} X-ray Observatory Center, which is operated by the Smithsonian Astrophysical Observatory for and on behalf of NASA under contract NAS8-03060. The Texas Advanced Computer Center at The University of Texas at Austin provided HPC resources under grant numbers TG-AST090105 and TG-AST130002, as did the PICSciE-OIT TIGRESS High Performance Computing Center and Visualization Laboratory at Princeton University. This work used the Extreme Science and Engineering Discovery Environment (XSEDE), which is supported by NSF grant OCI-1053575. MWK and AAS thank Merton College, Oxford and the Max-Planck/Princeton Center for Plasma Physics for travel support. This work benefitted from conversations with Ian Abel, Chris Chen, Geoffroy Lesur, Greg Hammett, Peter Porazik, Eliot Quataert, Francois Rincon, Prateek Sharma, and especially Steve Cowley.
\end{acknowledgments}

\bibliography{KSS14}

\end{document}